\documentclass[final,3p,times,twocolumn]{elsarticle}

\usepackage{graphicx}
\usepackage{amssymb}
\usepackage{units}
\usepackage{url}

\journal{Nuclear Instruments and Methods in Physics Research A}

\begin{document}

\begin{frontmatter}



\title{The $\gamma$-ray spectrometer HORUS and its applications for Nuclear Astrophysics}


\author[col]{L.~Netterdon\corref{cor1}}
\ead{lnetterdon@ikp.uni-koeln.de}
\cortext[cor1]{Corresponding author}
\author[col]{V.~Derya}
\author[col]{J.~Endres}
\author[col]{C.~Fransen}
\author[col]{A.~Hennig}
\author[col]{J.~Mayer}
\author[col]{C.~M\"uller-Gatermann}
\author[col]{A.~Sauerwein\fnref{fn1}}
\author[col]{P.~Scholz}
\author[col]{M.~Spieker}
\author[col]{A.~Zilges}

\address[col]{Institute for Nuclear Physics, University of Cologne, Z\"ulpicher Stra{\ss}e 77, D-50937 Cologne, Germany}

\fntext[fn1]{Present address: Institute for Applied Physics, Goethe University Frankfurt am Main, Germany}

\begin{abstract}
A dedicated setup for the in-beam measurement of absolute cross sections of astrophysically relevant charged-particle induced reactions is presented. These, usually very low, cross sections at energies of astrophysical interest are important to improve the modeling of the nucleosynthesis processes of heavy nuclei. Particular emphasis is put on the production of the $p$ nuclei during the astrophysical $\gamma$ process. The recently developed setup utilizes the high-efficiency $\gamma$-ray spectrometer HORUS, which is located at the \unit[10]{MV} FN tandem ion accelerator of the Institute for Nuclear Physics in Cologne. 

The design of this setup will be presented and results of the recently measured $^{89}$Y(p,$\gamma$)$^{90}$Zr reaction will be discussed. The excellent agreement with existing data shows, that the HORUS spectrometer is a powerful tool to determine total and partial cross sections using the in-beam method with high-purity germanium detectors. 
\end{abstract}


\end{frontmatter}


\section{Introduction}
\label{sec:introduction}

Reliable cross-section measurements of radiative capture-reactions are of utmost importance to understand the nucleosynthesis of nuclei heavier than iron, in particular the $\gamma$-process nucleosynthesis \cite{Rauscher13}. This process is believed to be mainly responsible for the production of the 30 - 35 neutron-deficient $p$ nuclei. Since the involved cross sections are typically in the $\mu$b range or even lower, a reliable measurement is very challenging. 

Various techniques are used for the measurement of radiative capture cross-sections. One of the most widely applied is the activation technique, which has provided a large amount of experimental data \cite{Yalcin09, Kiss11, Dillmann11, Sauerwein11, Halasz12, Netterdon13}. However, this technique is limited to reactions resulting in an unstable reaction product with appropriate half-lives. In order to overcome this limitation, mainly two different other methods are applied, namely the in-beam 4$\pi$-summing technique \cite{Tsagari04, Spyrou07} and the in-beam technique with high-purity germanium (HPGe) detectors \cite{Galanopoulos03,Sauerwein12, Harissopulos13}. The $\gamma$-ray spectrometer HORUS (\textbf{H}igh efficiency \textbf{O}bservatory for $\gamma$-\textbf{R}ay \textbf{U}nique \textbf{S}pectroscopy) \cite{Linnemann06} in combination with a recently developed target chamber can be used for the latter technique. With this method, the prompt $\gamma$-decays of the excited compound nucleus are observed. By measuring the angular distributions using a granular high-efficiency HPGe $\gamma$-ray detector array like HORUS at the University of Cologne, absolute reaction cross-sections can be determined. Moreover, since the individual $\gamma$-decay patterns are observed, further insight into the structure of the residual nucleus is provided. This might also lead to new nuclear structure results, $e.g.$ spin and parity assignments, see Ref. \cite{Sauerwein12}.

In this paper, a new setup utilizing the $\gamma$-ray spectrometer HORUS is presented. In Section~\ref{sec:horus} the design of the HORUS spectrometer is introduced followed by the target chamber dedicated for nuclear astrophysics experiments in Section~\ref{sec:chamber}. In Section~\ref{sec:experiments} the results of a first experiment on the $^{89}$Y(p,$\gamma$)$^{90}$Zr reaction are discussed. 

\section{The $\gamma$-ray spectrometer HORUS}
\label{sec:horus}

\begin{figure}[tb]
\centering
\includegraphics[width=0.8\columnwidth]{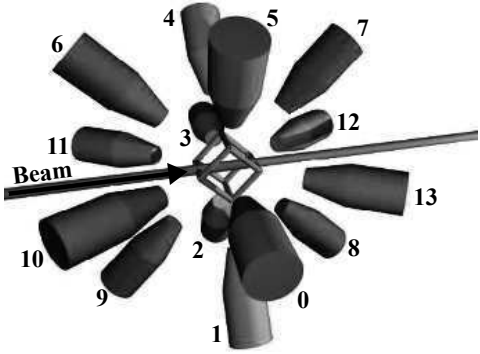}
\caption{Drawing of the HORUS $\gamma$-ray spectrometer. The 14 HPGe detectors are mounted on the eight cornes and six faces of a cube geometry. This allows the measurement at five different angles relative to the beam axis, namely at \unit[35]{$^\circ$} (detectors 12 and 13), \unit[45]{$^\circ$} (detectors 7 and 8), \unit[90]{$^\circ$} (detectors 0 to 5), \unit[135]{$^\circ$} (detectors 6 and 9), and \unit[145]{$^\circ$} (detectors 10 and 11).}
\label{fig:horus}
\end{figure}

The $\gamma$-ray spectrometer HORUS is located at the \unit[10]{MV} FN tandem ion accelerator at the Institute for Nuclear Physics in Cologne. It consists of up to 14 HPGe detectors, where six of them can be equipped with active BGO shields in order to actively suppress the Compton background. The HORUS spectrometer has a cubic geometry and the 14 detectors are mounted on its eight corners and six faces, respectively. Thus, the detectors are placed at five different angles with respect to the beam axis, namely at \unit[35]{$^\circ$}, \unit[45]{$^\circ$}, \unit[90]{$^\circ$},  \unit[135]{$^\circ$}, and \unit[145]{$^\circ$}, see Fig.~\ref{fig:horus}. This allows the measurement of angular distributions at five different angles, which is important to determine absolute cross sections. The distances of the HPGe detectors to the target can be freely adjusted. Due to the geometry of the target chamber and mounted BGO shields, the distances typically vary between \unit[9]{cm} and \unit[16]{cm}. Because of its flexible architecture, different types of HPGe can be mounted, $e.g.$, clover-type HPGe detectors or cluster-like detectors.

\subsection{Data acquisition}
The signal processing at HORUS is performed digitally using DGF-4C Rev. F modules, manufactured by the company XIA \cite{Hubbard-Nelson99, Skulski00}. Each module provides four input channels for the preamplifier signals of the semiconductor detectors as well as four channel-specific VETO inputs, which are used for the active Compton-background suppression with BGO shields. The preamplifier signals are digitized by ADCs with a depth of \unit[14]{bit} and a frequency of \unit[80]{MHz}. The modules allow the extraction of energy and time information as well as traces of the individual digitized preamplifier signals, if required by the experimentalist. An earlier revision of these modules has been successfully applied to the data acquisition of the Miniball spectrometer \cite{Warr13}.  Using these modules, it is possible to acquire data over a wide dynamic range of up to tens of MeV, which is important for astrophysical applications, since very high-energy $\gamma$ rays must be observed for this purpose. By storing the data event-by-event in a listmode format, it is possible to obtain $\gamma \gamma$ coincidence data, see Section \ref{sec:coincidence}. 

\subsection{Proton-energy determination}
\label{sec:energy}
\begin{figure}[tb]
\centering
\includegraphics[width=\columnwidth]{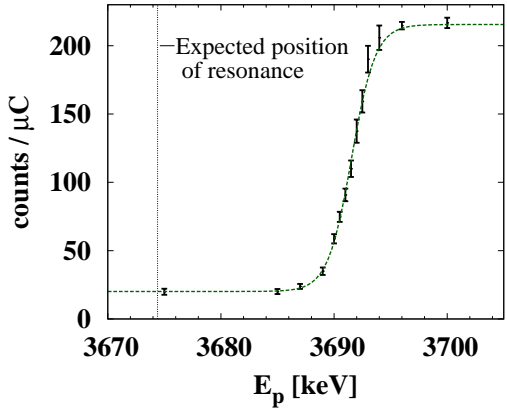}
\caption{(Color online) Resonance yield curve of the \unit[$E_p~=~3674.4$]{keV} resonance of the $^{27}$Al(p,$\gamma$)$^{28}$Si reaction. This resonance was scanned by varying the proton energy in small steps. This yield curve is used to exactly determine the energy of the incident protons as well as their energy spread. Moreover, this resonance is used for a relative full-energy peak efficiency calibration for $\gamma$-ray energies up to \unit[$E_\gamma \approx 10.5$]{MeV}, when measuring on top of the resonance. See text for details.}
\label{fig:yield_curve}
\end{figure}

In order to obtain a reliable determination of the proton energy, the \unit[$E_p=3674.4$]{keV} resonance of the $^{27}$Al(p,$\gamma$)$^{28}$Si reaction was used \cite{Brenneisen95}. By scanning this resonance, one obtains a calibration of the analyzing magnet. In a first step, the resonance was scanned by varying the projectile energy in small energy steps of down to \unit[0.5]{keV}. By normalizing the \unit[$E_\gamma=1779$]{keV} peak volume stemming from the $^{27}$Al(p,$\gamma$) reaction to the accumulated charge, a resonance yield curve was obtained, see Fig.~\ref{fig:yield_curve}. This yield curve shows a sharp low-energy edge and a plateau. The width of the sharp rising edge of the resonance yield curve is determined by the energy spread of the incoming protons. The location of the plateau is important for the efficiency calibration at high $\gamma$-ray energies, see Section~\ref{sec:efficiency}. Since the natural width of this resonance is smaller than \unit[2]{keV} \cite{Brenneisen95}, the width of the plateau is determined by the proton-energy loss inside the $^{27}$Al target. By fitting the resonance yield curve, a spread in the proton energy of \unit[$\pm$ 4]{keV} was found for the present setup. It is obvious from Fig.~\ref{fig:yield_curve}, that the center of the rising edge of the yield curve is not located at the literature value of \unit[$E_p=3674.4$]{keV}, but shifted by $\approx$~\unit[17]{keV} to higher energies. This has mainly two reasons. First, the earlier calibration of the analyzing magnet, relating the NMR measurement of the analyzing magnet to the particle energy, might not be exactly valid, since this is very sensitive to the exact geometry of the beamlines and slits defining the entrance to the analyzing magnet. For this purpose, a measurement of the exact proton energy must be done during every conducted experiment, where the particle energy must be precisely known. Secondly, a finite opening angle of the slits in front of the analyzing magnet allows a geometrical uncertainty for the beam entrance into the analyzing magnet, $i.e.$ a skew pathway of the beam. However, since these parameters do not change during the experiment, the beam uncertainty and energy offset can be considered as constant. This constant offset must be taken into account in the data analysis, especially when it comes to determine the energy straggling inside the target material, or when comparing the measured cross-section results to theoretical calculations.

\subsection{Measurement of the full-energy peak efficiency}
\label{sec:efficiency}
\begin{figure}[tb]
\centering
 \includegraphics[width=\columnwidth]{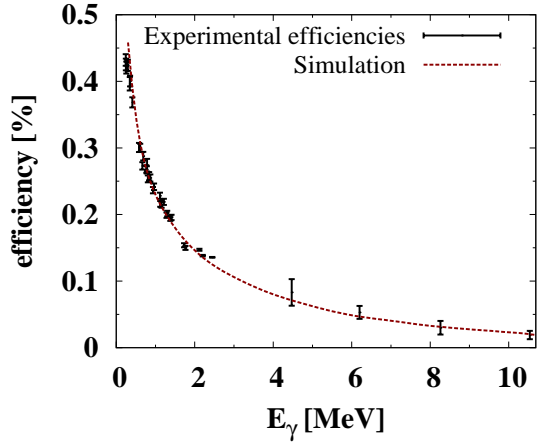}
\caption{(Color online) Full-energy peak efficiency for one of the HPGe detectors with a distance of \unit[14]{cm} to the target. The efficiencies for $\gamma$-ray energies higher than \unit[$E_\gamma = 4$]{MeV} were obtained by using the \unit[$E_p=3674.4$]{keV} resonance of the $^{27}$Al(p,$\gamma$)$^{28}$Si reaction. The decay of this resonance provides $\gamma$ rays with energies of up to \unit[$E_\gamma~=~10509$]{keV}, which are used for a relative efficiency determination. The remaining ones were obtained using calibrated radioactive sources. The experimental full-energy peak efficiencies are compared to the efficiencies obtained with a \textsc{Geant4} simulation (dashed line), which shows an excellent agreement over an energy range from \unit[$E_\gamma \approx 350$]{keV} up to the highest $\gamma$-ray energy of \unit[$E_\gamma \approx 10500$]{keV}.} 
\label{fig:efficiency}
\end{figure}

The absolute full-energy peak efficiency is mandatory to derive absolute reaction cross-sections. Up to a $\gamma$-ray energy of about \unit[3.6]{MeV} the full-energy peak efficiency can be determined using standard calibrated radioactive sources. However, for astrophysical applications, the absolute full-energy peak efficiencies for $\gamma$-ray energies of up to more than \unit[10]{MeV} must be precisely known. For this purpose, more sophisticated techniques such as measuring resonances of capture reactions or inelastic scattering reactions must be applied. For this setup, the \unit[$E_p=3674.4$]{keV} resonance of the $^{27}$Al(p,$\gamma$)$^{28}$Si reaction was used \cite{Brenneisen95}, which allows the determination of the full-energy peak efficiency up to an energy of \unit[$E_\gamma=10509$]{keV}.
The resonance with an excitation energy of \unit[$E_x=15127$]{keV} is populated, when measuring on top of the resonance plateau, see Fig.~\ref{fig:yield_curve}. 
The absolute branching ratios of the decay of this resonant state are known \cite{Brenneisen95}. They can be used to determine the relative full-energy peak efficiencies, which are subsequently scaled to an absolute efficiency calibration obtained with calibration sources. For the present case, the $\gamma$ rays with energies of \unit[$E_\gamma=$10509]{keV}, \unit[8239]{keV}, \unit[6182]{keV}, and \unit[4458]{keV} were used for the efficiency calibration. Fig.~\ref{fig:efficiency} shows the absolute full-energy peak efficiency for a typical HPGe detector used in the HORUS spectrometer as a function of $\gamma$-ray energy. The full-energy peak efficiencies for the lower energy range up to a $\gamma$-ray energy of \unit[$E_\gamma \approx 2.5$]{MeV} were obtained using standard calibrated $^{152}$Eu and $^{226}$Ra sources. Finally, the full-energy peak efficiencies were obtained by fitting a function of the form 

\begin{equation}
f(E_\gamma) = a \cdot \exp \left(b\cdot E_\gamma\right) + c \cdot \exp \left(d \cdot E_\gamma\right)
\end{equation}
to the experimental efficiencies.

In order to obtain a reliable efficiency determination over the whole energy region and allow for coincidence summing effects, Monte Carlo simulations were performed using the \textsc{Geant4} toolkit \cite{Geant4}. These simulations agree very well with the experimentally obtained full-energy peak efficiencies over an energy range from \unit[$E_\gamma \approx 350$]{keV} up to the highest $\gamma$-ray energy of \unit[$E_\gamma \approx 10500$]{keV}, see Fig.~\ref{fig:efficiency}. Due to the large distance of the HPGe detector to the target and low count rates, summing effects are negligible. Below a $\gamma$-ray energy of \unit[$E_\gamma~\approx~350$]{keV}, the \textsc{Geant4} simulation tends to overestimate the experimental data. This is due to the fact, that the very details of the detector geometry become significant at such low energies. This includes, $e.g.$, the exact thickness of the detector end caps or dead layers of the HPGe detectors, which might not be known to a sufficient precision. However, this does not play a significant role for the efficiency determination, since the full-energy peak efficiencies for the lower energies can be well determined using standard calibration sources. 

\subsection{$\gamma \gamma$ coincidences}
\label{sec:coincidence}
The combination of the high granularity of the HORUS spectrometer and the event-by-event data format facilitates the construction of $\gamma \gamma$ coincidence matrices. The $\gamma \gamma$ coincidence technique is a powerful tool to suppress the beam-induced background. Although it cannot be applied directly to the determination of absolute cross sections, it is most helpful to unambigiously identify the $\gamma$-ray transitions visible in the spectra. With this, it can be proved, for instance, that the observed $\gamma$-ray transition corresponding to the reaction of interest is free from contaminants from other decaying nuclei resulting from target contaminants. 

\begin{figure}[tb]
\centering
\includegraphics[width=\columnwidth]{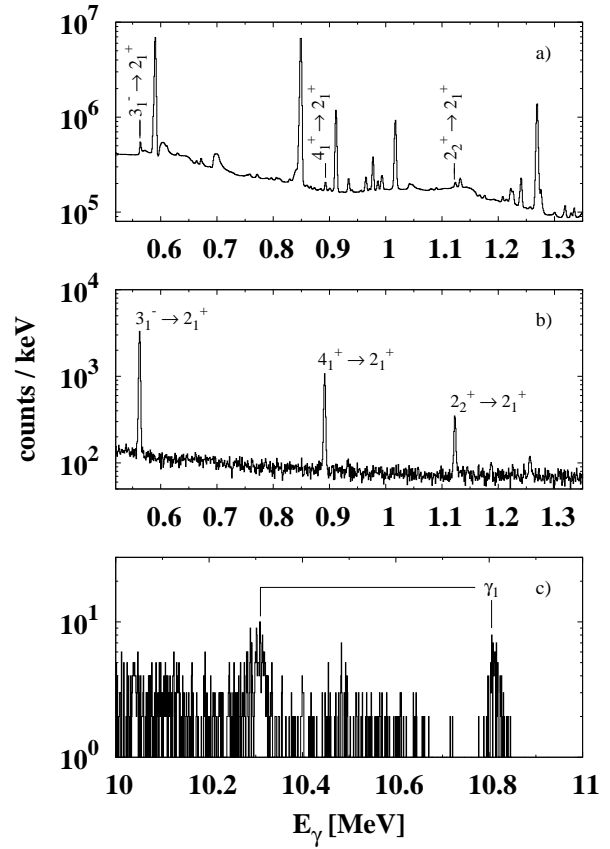}
\caption{Excerpt from a $\gamma \gamma$ coincidence spectrum of the $^{89}$Y(p,$\gamma$)$^{90}$Zr reaction using \unit[4.7]{MeV} protons. The upper panel a) shows a singles spectrum, $i.e.$, no gate was applied. The three marked transitions are hardly visible due to the beam-induced background. Afterwards, a gate was set on the $\gamma$-ray transition from the first excited $J^\pi=2_1^+$ state to the ground state. The low-energy part b) shows the transitions from higher lying discrete states in $^{90}$Zr feeding the $J^\pi=2_1^+$ state. In the high-energy part c) one can clearly recognize the de-excitation from the compound state to the first excited state, denoted as $\gamma_1$, together with its single escape peak.} 
\label{fig:coincidence}
\end{figure}

Figs.~\ref{fig:coincidence} a) to c) demonstrate the $\gamma \gamma$ coincidence method. Fig.~\ref{fig:coincidence} a) shows a part of a $\gamma$-ray spectrum of the $^{89}$Y(p,$\gamma$)$^{90}$Zr reaction using \unit[4.7]{MeV} protons, where no gate was applied. The three marked transitions from higher lying states feeding the \unit[2186]{keV}-state in $^{90}$Zr can hardly be recognized due to the large beam-induced background. After a gate on the $\gamma$-ray transition from the first excited $J^\pi=2_1^+$ state to the ground state was applied, these transitions become clearly visible, see Fig.~\ref{fig:coincidence} b). Fig.~\ref{fig:coincidence} c) shows the high-energy part of the coincidence spectrum. The transition from the excited compound state to the first excited state, denoted as $\gamma_1$, can be clearly identified, together with its single escape peak. 

Moreover, $\gamma$-ray transitions from the observed compound nucleus with a low intensity might vanish in the beam-induced background. These transitions would be missing later on, when the total cross section is determined. Using the $\gamma \gamma$ coincidence technique, the beam-induced background can be reduced so efficiently, that even the weakest $\gamma$-ray transitions become visible in the coincidence spectra. The absolute influence of such a transition on the cross section might not be determined using $\gamma \gamma$ coincidences. But it strongly supports determining an upper limit of the impact of a transition on the total cross section, which is hidden in the beam-induced background. Thus, using this method, systematic uncertainties concerning missing $\gamma$-strength during the data analysis are drastically reduced.

\section{Target chamber design}
\label{sec:chamber}
\begin{figure}[tb]
\centering
 \includegraphics[width=\columnwidth]{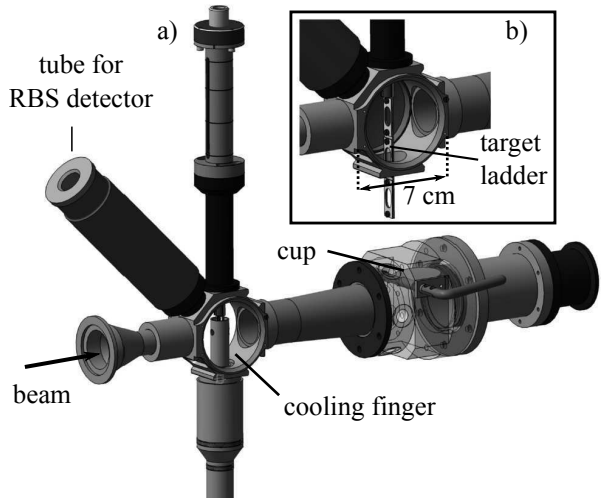}
\caption{Illustration of the target chamber used for nuclear astrophysics experiments \cite{Thiel12}. A cooling finger around the target ladder prevents residual gas deposits on the target. The inset b) shows the inside of the target chamber without the cooling finger. A silicon detector for RBS measurements is also available during the experiments, see Section~\ref{sec:rbs}. The chamber is coated with tantalum to avoid competetive reactions on the aluminum housing. The cup closely behind the target can be tilted to stop the beam downstream the HORUS spectrometer.} 
\label{fig:targetchamber}
\end{figure}

The target chamber mounted inside the HORUS spectrometer was optimized for nuclear astrophysics experiments. The chamber itself measures \unit[7]{cm} in length and \unit[5.5]{cm} in width, see Fig.~\ref{fig:targetchamber}. The dimensions of the target chamber were kept as small as possible, in order to facilitate a preferably short distance from the HPGe detectors to the target, which allows a high full-energy peak efficiency. The body of the target chamber is made of aluminum with a thickness of \unit[2]{mm}. It provides a tantalum coating inside with a thickness of \unit[0.1]{mm}, which is used to suppress competitive reactions on the aluminum housing. The thickness of the tantalum coating was optimized in a way, that on the one hand it is thick enough to stop \unit[5]{MeV} protons and \unit[15]{MeV} $\alpha$-particles, but thin enough not to significantly absorb $\gamma$ rays down to an energy of \unit[$E_\gamma=250$]{keV}. Moreover, this coating can be removed for activation experiments, when $\gamma$ rays with lower energies have to be observed with the HORUS spectrometer after the activation.

The target is surrounded by a copper tube cooled by $\mathrm{LN}_2$. This tube serves as a cooling finger and is used to minimize residual gas deposits on the target material. Fig.~\ref{fig:targetchamber} a) shows an overview of the target chamber including the cooling finger around the target ladder. The inset b) shows a close-up view of the chamber with the cooling finger removed, offering a view on the target ladder.  

\subsection{Current read-out}
Since the total number of particles impinging on the target must be known for the cross-section determination, the beam current is read out at three different positions. Firstly, the current is measured at the target itself and at a Faraday cup, which is located at a distance of \unit[15]{cm} behind the target. This Faraday cup can be tilted, $i.e.$, one can choose to stop the beam in the Faraday cup closely behind the target or the beam dump downstream the HORUS spectrometer. Additionally, the current is separately read out at the target chamber itself, in order to measure released secondary electrons as well as scattered beam particles. The accumulated charge is individually determined by current integrators with an overall uncertainty of about \unit[5]{\%}. A negatively charged aperture with a voltage of \unit[$U_S=-400$]{V} at the entrance of the target chamber is used to prevent secondary electrons from hitting the beamline upstream.

\subsection{RBS setup}
\label{sec:rbs}
\begin{figure}[tb]
\centering
\includegraphics[width=\columnwidth]{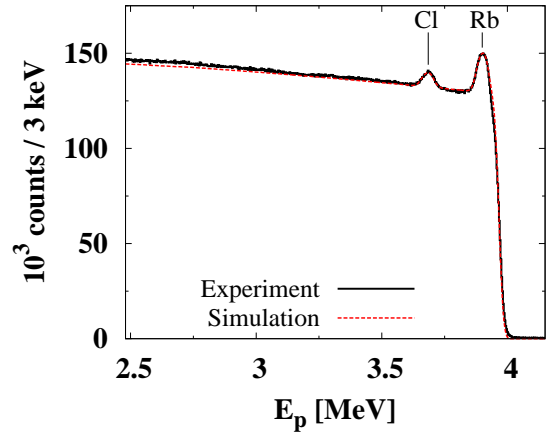}
\caption{(Color online) Relevant part of the RBS spectrum of a measured RbCl target, using the built-in RBS detector with \unit[4]{MeV} protons. The spectrum is very well reproduced by a simulation using the SIMNRA code (dashed line). The target thickness was additionally measured at the RUBION facility in Bochum, Germany, prior to the experiment. Both measurement yield an excellent agreement within the experimental uncertainties. The peaks belonging to Rb (right) and Cl (left) can be clearly identified on top of the  Au backing. See text for details.} 
\label{fig:rbs}
\end{figure}

The target chamber additionally houses a silicon detector, which is used for Rutherford Backscattering Spectrometry (RBS) measurements, see Fig~\ref{fig:targetchamber} a). It is placed at an angle of \unit[135]{$^\circ$} relative to the beam axis at a distance of \unit[11]{cm} to the target to monitor target stability during the experiment and to measure the target thickness. Since the RBS setup became available after the $^{89}$Y(p,$\gamma$) measurement, the proof of principle for this setup is shown for a measurement on the $^{85}$Rb(p,$\gamma$) reaction in the following.
 A typical RBS spectrum of an $^{85}$Rb target irradiated with protons with an energy of \unit[$E_p=4$]{MeV} is shown in Fig.~\ref{fig:rbs}. This target was prepared by evaporating RbCl, enriched to \unit[(99.78$\pm$0.02)]{\%} in $^{85}$Rb, onto a \unit[150]{$\frac{\mathrm{mg}}{\mathrm{cm}^2}$} thick gold backing. The two peaks belonging to Rb (right) and Cl (left) can be clearly identified. The measured RBS spectrum was simulated using the SIMNRA code \cite{Mayer02}, which yields a very good agreement, see Fig.~\ref{fig:rbs}. 

The thickness of this target was measured prior to the experiment at the RBS facility at the RUBION dynamitron-tandem accelerator at the Ruhr-Universit\"at Bochum and amounts to an areal particle density of Rb atoms of \unit[$(2.01 \pm 0.09) \times 10^{18}$]{$\frac{1}{\mathrm{cm}^2}$}. The RBS measurement in Cologne yields \unit[$(1.97 \pm 0.16) \times 10^{18}$]{$\frac{1}{\mathrm{cm}^2}$}, which is in excellent agreement with the Bochum results. Hence, one can conclude, that no target material was lost during the irradiation.

\section{Cross-section measurement of the $^{89}$Y(p,$\gamma$)$^{90}$Zr reaction}
\label{sec:experiments}
The astrophysically relevant $^{89}$Y(p,$\gamma$)$^{90}$Zr reaction was chosen as a test case for the recently developed experimental setup in Cologne. This nucleus is located in a mass region, where the $p$-nuclei abundances are not well reproduced by reaction-network calculations \cite{Woosley07}. It is well-suited as a test case, since it was recently measured in two experiments using the in-beam method with HPGe detectors \cite{Harissopulos13} as well as the 4$\pi$-summing technique \cite{Tsagari04}. 

The reaction cross-section was measured at five different proton energies ranging from \unit[$E_p~=~3.7$]{MeV} to \unit[$E_p~=~4.7$]{MeV}. For this experiment, a natural Y target containing \unit[99.9]{\%} of $^{89}$Y was used. It was prepared by vacuum evaporation on a \unit[130]{$\frac{\mathrm{mg}}{\mathrm{cm}^2}$} thick tantalum backing, where the beam was stopped. The target had a thickness of \unit[(583 $\pm$ 24)]{$\frac{\mu\mathrm{g}}{\mathrm{cm}^2}$}. An RBS measurement at the Ruhr-Universit\"at Bochum before and after the measurement ensured, that no target material was lost during the irradiation.

\begin{figure}[tb]
\centering
\includegraphics[width=\columnwidth]{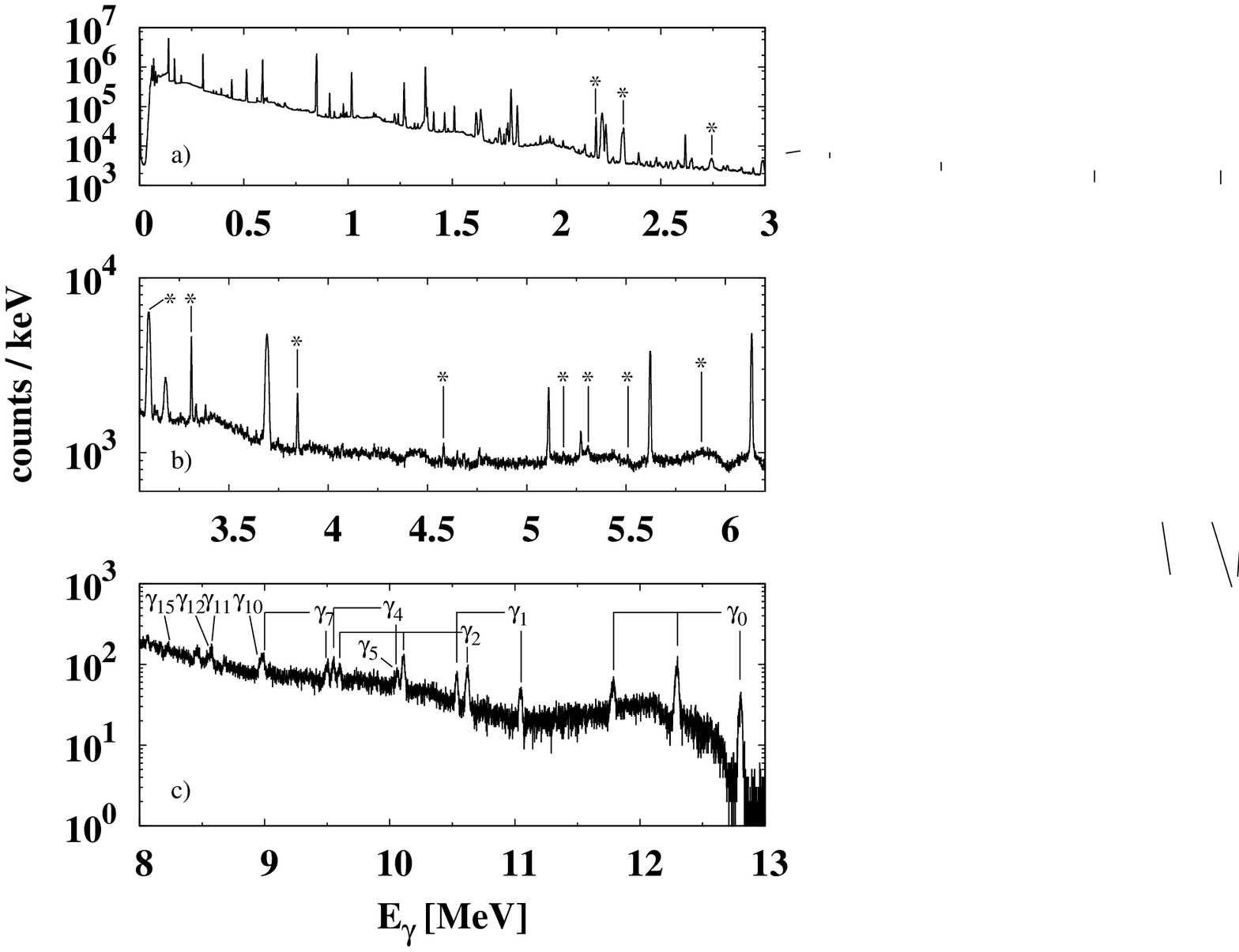}
\caption{Typical $\gamma$-ray spectrum recorded during the irradiation of $^{89}$Y with \unit[4.5]{MeV} protons. This spectrum was obtained by summing over all HPGe detectors at an angle of \unit[90]{$^{\circ}$} relative to the beam axis. All transitions to the ground state of the reaction product $^{90}$Zr are marked with an asterisk. The high-energy part c) shows de-excitations from the compound state to the ground state or one of the excited states. The transition to the ground state is indicated by $\gamma_0$, to the first excited state as $\gamma_1$, and so on. De-excitations up to the $15^{th}$ excited state were observed. For the strongest transitions, the single and / or double escape peak is marked, too.}
\label{fig:89Y_spectrum}
\end{figure}

The target was bombarded for several hours with beam currents ranging from \unit[1]{nA} to \unit[60]{nA}. The large span of the beam current was due to technical limitations of the accelerator. A typical $\gamma$-ray spectrum for the $^{89}$Y(p,$\gamma$) reaction using a proton energy of \unit[$E_p~=~4.5$]{MeV} is shown in Figs.~\ref{fig:89Y_spectrum} a) - c). It was obtained by summing up the $\gamma$-ray spectra of the six HPGe detectors positioned at an angle of \unit[90]{$^\circ$} relative to the beam axis. Despite the rather strong beam-induced background, all relevant $\gamma$-ray transitions populating the ground state in the reaction product $^{90}$Zr can be clearly identified, marked by asterisks in Figs.~\ref{fig:89Y_spectrum} a) and b). Moreover, de-excitations of the compound state up to the $15^{th}$ excited state, denoted as $\gamma_i$, can be observed, see Fig.~\ref{fig:89Y_spectrum} c). The excitation energies were adopted from Ref.~\cite{NNDC}.

\subsection{Cross-section determination} 
\begin{figure}[tb]
\centering
 \includegraphics[width=\columnwidth]{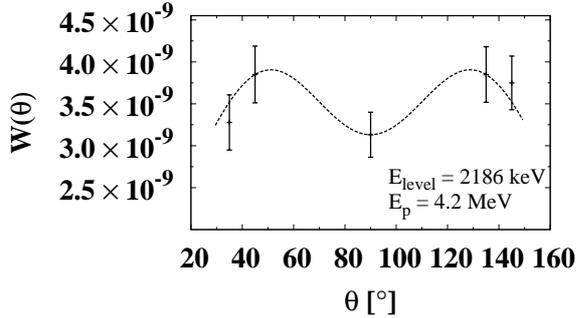}
\caption{Angular distribution for the $2_1^+ \rightarrow 0_{\mathrm{g.s.}}^+$ $\gamma$-ray transition in $^{90}$Zr. The incident proton energy was \unit[$E_p=4.2$]{MeV}. The dashed line depicts the sum of Legendre polynomials fitted to the experimental data.} 
\label{fig:angulardistribution}
\end{figure}

In order to determine the total cross section $\sigma$ of the (p,$\gamma$) reaction, the number of produced compound nuclei $N_{\mathrm{comp}}$ must be known, which is given by
\begin{equation}
\label{eq:cross-section}
N_{\mathrm{comp}} = \sigma \cdot N_{\mathrm{proj}} \cdot m_{\mathrm{target}} ,
\end{equation}
where $N_{\mathrm{proj}}$ is the number of projectiles, and $m_{\mathrm{target}}$ is the areal particle density of target nuclei. $N_{\mathrm{comp}}$ is derived by measuring the angular distributions of all $\gamma$ rays populating the ground state. The measured intensities of these $\gamma$ rays are then corrected for the respective number of impinging projectiles for each beam energy, the full-energy peak efficiency as well as the dead time of the data-acquisition system. Subsequently, for every $\gamma$-ray transition a sum of Legendre polynomials was fitted to the experimental angular distributions: 

\begin{equation}
W^i\left(\theta\right) = A_0^i \left(1+\sum_{k=2,4} \alpha_k P_k(\cos \theta)\right) ,
\end{equation}
with the energy-dependent coefficients $A_0$ and $\alpha_k~(k~=~2,4)$. 
An example of an angular distribution for the $\gamma$-ray transition from the \unit[$E_x = 2186$]{keV} \cite{NNDC} level to the ground state in $^{90}$Zr for an incident proton energy of \unit[$E_p=4.2$]{MeV} is shown in Fig. \ref{fig:angulardistribution}. The cross section can be calculated from the absolute coefficients of the respective angular distributions $A_0^i$:
\begin{equation}
\sigma = \frac{\sum_{i=1}^N A_0^i}{m_{\mathrm{target}}} ,
\end{equation}
where $N$ is the total number of coefficients, $i.e.$, the number of considered $\gamma$-ray transitions. Further details about the data-analysis procedure can be found, $e.g.$, in Ref.~\cite{Harissopulos13}. 

In the compound nucleus $^{90}$Zr, there is an isomeric $J^\pi~=~5^-$ state at an excitation energy of \unit[$E_x~=~2319$]{keV} with a half-life of \unit[809.2]{ms} \cite{NNDC}. Thus, one must take care, that no yield stemming from this transition is lost, if only the prompt $\gamma$-rays are detected. This was accomplished by acquiring data until \unit[$\approx 5$]{s} after the irradiation was stopped, in order to guarantee that all $\gamma$-rays stemming from this ground-state transition are detected.

By using the method of in-beam $\gamma$-ray spectroscopy with HPGe detectors, it is also possible to observe de-excitations of the compound nucleus to various excited states. In the cases of activation or 4$\pi$-summing techniques, this is only feasible for isomeric states with a sufficiently long half-life. By investigating the angular distributions for each of these $\gamma$-ray transitions, it is possible to derive partial cross sections. The possibility to measure partial cross sections is a tremendous advantage compared to the other aforementioned experimental techniques. Partial cross sections can be used to experimentally constrain the $\gamma$-strength function, which is an important input parameter for theoretically predicted astrophysical reaction rates, not only for $\gamma$-process calculations, but also for, $e.g.$, neutron-capture reactions during the $r$ process \cite{Arnould07}.

\subsection{Experimental results}
\begin{figure}[thb]
\centering
 \includegraphics[width=\columnwidth]{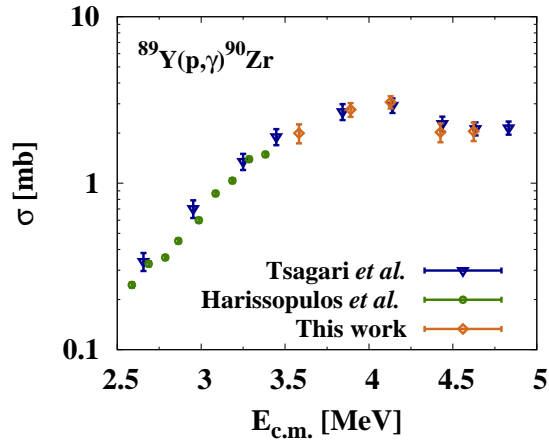}
\caption{(Color online) Total cross section for the $^{89}$Y(p,$\gamma$)$^{90}$Zr reaction as a function of the effective center-of-mass energy $E_{\mathrm{c.m.}}$. The results are compared to previous measurements using the 4$\pi$-summing technique (triangles) \cite{Tsagari04} and the in-beam method with HPGe detectors (cirlces) \cite{Harissopulos13}. The present data are found to be in excellent agreement with the previous measurements.} 
\label{fig:results}
\end{figure}

The experimental total cross sections of the $^{89}$Y(p,$\gamma$)$^{90}$Zr reaction obtained in this work are given in Table~\ref{tab:results} and shown in Fig.~\ref{fig:results}. The effective energies given in the first column of Table~\ref{tab:results} were obtained from
\begin{equation}
E_p = E_0 - \frac{\Delta E}{2},
\end{equation}
where $E_0$ is the incident proton energy including the aforementioned \unit[17]{keV} offset, see Section~\ref{sec:energy}, and $\Delta E$ is the average energy loss inside the target material, that was obtained using the \textsc{Srim} code \cite{Srim12} and amounts to \unit[24-28]{keV} depending on the incident energy. The energy straggling inside the target material was \unit[$\approx 8$]{keV} for all energies.

\begin{table}[tb]
\centering
\caption{Total cross sections $\sigma$ measured for the $^{89}$Y(p,$\gamma$)$^{90}$Zr reaction as a function of the effective center-of-mass energy $E_{\mathrm{c.m.}}$.}
\begin{tabular}{c c}
$E_{\mathrm{c.m.}}$ [keV]	&	$\sigma$ [mb] \\ \hline
\vspace{-0.2cm}
 & \\
3583 $\pm$ 8	& 1.99  $\pm$ 0.27	\\
3891 $\pm$ 8	& 2.77  $\pm$ 0.30	\\
4129 $\pm$ 8	& 3.07 $\pm$ 0.27	\\
4426 $\pm$ 8	& 2.02 $\pm$ 0.25 	\\
4624 $\pm$ 8	& 2.05 $\pm$ 0.26	\\ \hline

\end{tabular}
\label{tab:results}
\end{table}

The uncertainties in the measured cross sections are composed of the uncertainties in target thickness (\unit[$\approx~4$]{\%}), accumulated charge (\unit[$\approx~5$]{\%}), detector efficiency (\unit[$\approx~8$]{\%}), and the statistical error of the fit of the angular distribution (\unit[$\approx~4 - 7$]{\%}).

A comparison of the experimental cross sections to previously measured data of Refs.~\cite{Tsagari04,Harissopulos13} is given in Fig.~\ref{fig:results}. The present results are in excellent agreement with the previous data. Hence, one can conclude, that the recently developed setup at the Institute for Nuclear Physics in Cologne has become fully operational for cross-section measurements.
Note, that the measured partial cross sections are not discussed here. The discussion and their possible impact on the $\gamma$-strength function will be subject of a forthcoming publication. 

\section{Conclusions}
\label{sec:conclusions}
In this article, the recently developed setup for cross-section measurements relevant for nuclear astrophysics utilizing the high-efficiency $\gamma$-ray detector array HORUS was presented. HORUS is a highly flexible spectrometer, where also other detector types such as cluster-like detectors or clover-type HPGe detectors can be mounted. It is possible to obtain $\gamma \gamma$ coincidence data, which is a powerful tool to suppress the beam-induced background and to clearly identify the $\gamma$-ray transitions of interest. This method drastically reduces systematic uncertainties concerning missing weak $\gamma$-ray transitions in the spectra.

The recently developed setup for nuclear astrophysics experiments  was used to determine total cross sections of the $^{89}$Y(p,$\gamma$)$^{90}$Zr reaction at five different energies between \unit[$E_p=3.7 - 4.7$]{MeV}. The data at hand is compared to previously measured data of Refs.~\cite{Tsagari04,Harissopulos13} and an excellent agreement is found. 
Moreover, for the first time partial cross sections of this reaction were measured. The results and their discussion concerning the impact of these measurements on the $\gamma$-strength function will be presented in a forthcoming publication.
The method of in-beam $\gamma$-ray spectroscopy allows studying reactions with a stable reaction product. Thus, the experimental data needed for, $e.g.$, $\gamma$-process network calculations can be widely extended. The $\gamma$-ray spectrometer HORUS embodies an excellent tool to study charged-particle induced reactions at energies of astrophysical interest. 

\section*{Acknowledgments}
The authors thank A. Dewald and the accelerator staff at the Institute for Nuclear Physics in Cologne for providing excellent beams. Moreover, we gratefully acknowledge the assistance of H.-W. Becker and D. Rogalla of the Ruhr-Universit\"at Bochum on RBS measurements. This project was partly supported by the Deutsche Forschungsgemeinschaft under contract ZI 510/5-1.

\end{document}